\documentclass[12pt]{article}
\usepackage{epsf}
\usepackage{amsmath,amssymb}
\usepackage{pdfpages}
\usepackage{graphicx}
\usepackage{comment}
\usepackage[driverfallback=dvipdfm]{hyperref}

\begin{document}

\newcommand{\lsim}{\stackrel{<}{_\sim}}
\newcommand{\gsim}{\stackrel{>}{_\sim}}

\newcommand{\rem}[1]{{$\spadesuit$\bf #1$\spadesuit$}}

\renewcommand{\theequation}{\thesection.\arabic{equation}}

\renewcommand{\thefootnote}{\fnsymbol{footnote}}
\setcounter{footnote}{0}

\begin{titlepage}

\def\thefootnote{\fnsymbol{footnote}}

\begin{center}

\hfill UT-19-29\\
\hfill December, 2019\\

\vskip .75in

{\Large \bf

  Determining Wino Lifetime in Supersymmetric Model\\
  at Future 100 TeV $pp$ Colliders \\

}

\vskip .75in

{\large So Chigusa${}^{a}$, Yusuke Hosomi${}^{a}$, Takeo Moroi${}^{a}$ and Masahiko Saito${}^{b}$}

\vskip 0.25in

{\em ${}^{a}$Department of Physics, University of Tokyo,
Tokyo 113-0033, Japan \\
${}^{b}$International Center for Elementary Particle Physics,
University of Tokyo, \\ Tokyo 113-0033, Japan}

\end{center}
\vskip .5in

\begin{abstract}

  We discuss a possibility to measure the lifetime of charged Wino in
  supersymmetric model at future $100\ {\rm TeV}$ $pp$ colliders,
  assuming that (neutral) Wino is the lightest superparticle (LSP).
  In the Wino LSP scenario, the charged Wino has a lifetime of about
  $0.2\ {\rm ns}$, and its track may be reconstructed in particular by
  the inner pixel detectors.  We show that the lifetime of charged
  Wino may be measured by using the information about the distribution
  of the flight lengths of charged Winos.  We propose a procedure for
  the lifetime determination and show how the accuracy changes as we
  vary the mass spectrum of superparticle.  We also discuss the
  effects of the detector layouts on the lifetime determination.

\end{abstract}

\end{titlepage}

\renewcommand{\thepage}{\arabic{page}}
\setcounter{page}{1}
\renewcommand{\thefootnote}{\#\arabic{footnote}}
\setcounter{footnote}{0}

\section{Introduction}
\label{sec:intro}
\setcounter{equation}{0}

Low energy supersymmetry (SUSY), with superparticles at the mass scale
much lower than the Planck scale, has been attracted attentions.  Even
though no direct evidence of superparticles has been experimentally
found yet, it is still a well-motivated candidate of physics beyond
the standard model. In particular, in models with low energy
supersymmetry, gauge coupling unification at the scale of grand
unified theory (GUT), i.e., $\sim 10^{16}\ {\rm GeV}$, is possible.
In addition, more importantly for our study, the lightest
superparticle (LSP) in SUSY model with $R$-parity conservation can be
dark matter.  The SUSY dark matter is an important target not only of
direct detection experiments but also of high energy colliders.
Notably, the collider phenomenology of SUSY dark matter depends on the
properties of the LSP.

In the present study, we concentrate on SUSY model in which Winos,
which are superpartners of $SU(2)_L$ gauge bosons, are lighter than
other superparticles and discuss its collider phenomenology.  This
class of model is motivated by the present constraints on low energy
SUSY.  First, the neutral Wino can be a viable candidate of dark
matter.  In addition, the Wino LSP naturally shows up from so-called
minimal gravity mediation model \cite{Ibe:2006de, Ibe:2011aa,
  ArkaniHamed:2012gw} based on anomaly mediation \cite{Randall:1998uk,
  Giudice:1998xp}.  In such a model, masses of gauginos are of the
order of $(0.1-1)\ {\rm TeV}$, while those of other superparticles are
a few orders of magnitude heavier.  Notably, such a mass spectrum is
well motivated from the Higgs mass point of view because heavier
superparticles, in particular heavy stops, can push up the Higgs mass
to the observed value of about $125\ {\rm GeV}$ via radiative
corrections \cite{Okada:1990vk, Haber:1990aw, Ellis:1990nz,
  Ellis:1991zd}.

In the SUSY model of our interest, the primary targets of the collider
study are Winos as well as other gauginos.  The thermal relic
abundance of Wino becomes equal to the dark matter density if its mass
is about $2.9\ {\rm TeV}$ \cite{Hisano:2006nn}, while lighter Wino can
also become dark matter if Winos are non-thermally produced in the
early universe \cite{Giudice:1998xp, Moroi:1999zb}.  Combined with the
present collider bounds on superparticles, it may be the case that the
Winos (and other superparticles) are out of the kinematic reach of the
LHC experiment.  Such a possibility motivates us to consider more
energetic colliders than the LHC.  In particular, $pp$ colliders with
the center of mass energy of $\sim 100\ {\rm TeV}$, called future
circular collider (or FCC-hh), is now seriously discussed.

Here, we consider the collider phenomenology of supersymmetric model
with Wino LSP at FCC-hh.  In the previous studies, it has been
discussed that the discovery \cite{Saito:2019rtg} and the mass
measurements \cite{Asai:2019wst} of gauginos are possible at FCC-hh,
particularly relying on the existence of disappearing tracks of
charged Winos.\footnote
{We may also use the study of mono-jet events \cite{Han:2018wus} and
  precision study of the Drell-yan processes \cite{Chigusa:2018vxz,
    DiLuzio:2018jwd, Matsumoto:2018ioi, Abe:2019egv} for the discovery
  of the signals of electroweakly interacting particles, like Winos,
  at the FCC-hh.}
In the case of Wino LSP, it is often the case that the mass difference
between charged and neutral Winos is induced dominantly by the
radiative correction due to the electroweak gauge bosons; the charged
Wino becomes slightly heavier than the neutral one and the mass
difference is given by $\sim 160\ {\rm MeV}$.  As a result, the
charged Wino becomes fairly long-lived; its lifetime is often $\sim
0.2\ {\rm ns}$ and is insensitive to the mass spectrum of
superparticles (as far as superparticles other than gauginos are much
heavier than Wino).  Then, once produced at the colliders, the charged
Wino may fly $O(1-10)\ {\rm cm}$ and may be identified as a short high
$p_T$ track.  In order to understand the properties of Winos,
measurement of the lifetime of charged Wino is an important step.

In this letter, we study the possibility of measuring the lifetime of
charged Wino at FCC-hh.  If the Wino mass is less than $\sim 2.9\ {\rm
  TeV}$, which is the upper bound on the Wino mass from the point of
view of dark matter, it is expected that the charged Wino is within
the reach of FCC-hh \cite{Saito:2019rtg} and that the Wino mass can be
determined \cite{Asai:2019wst}.  Here, using SUSY events with charged
Wino production, we show that the lifetime of the charged Wino can be
also determined.  We discuss the basic procedure to determine the
charged Wino lifetime at FCC-hh, and show the expected accuracy of the
lifetime measurement.  The organization of this letter is as follows.
In Section \ref{sec:formalism}, we explain the method of the lifetime
measurements at the FCC-hh.  We also summarize important features of
the model of our interest.  Then, in Section \ref{sec:mc}, we show our
numerical results.  Section \ref{sec:conclusion} is devoted for
conclusions and discussion.

\section{Formalism}
\label{sec:formalism}
\setcounter{equation}{0}

Let us explain the setup of our analysis.  Here, we concentrate on
supersymmetric models in which Winos are lighter than other
supersymmetric particles.  Detailed mass spectrum of superparticles
for our Monte Carlo~(MC) analysis will be explained in the next
section.  We also assume that the mass difference between the charged
and neutral Winos dominantly comes from radiative effects due to
electroweak gauge bosons.  In such models, neutral Wino becomes
slightly lighter than charged ones, and hence neutral Wino becomes the
LSP while charged Wino becomes long-lived.  The mass difference is
predicted to be $\sim 160\ {\rm MeV}$ which gives the lifetime of the
charged Wino of $\sim 0.2\ {\rm ns}$ \cite{Cheng:1998hc, Feng:1999fu,
  Gherghetta:1999sw, Yamada:2009ve, Ibe:2012sx}.  Here, we take the
canonical lifetime of charged Wino to be $c\tau=5.75\ {\rm cm}$
\cite{Ibe:2012sx}, with $c$ being the speed of light.  In the
following, we study how well we can determine the lifetime of charged
Wino at the FCC-hh experiment. Although our primary interest is to
determine the Wino lifetime, we vary the input value of the lifetime
to see how the sensitivity depends on it.

Once produced, charged Wino travels finite distance and decays into
neutral Wino (and charged pion).  In particular, some of charged Winos
travel long enough to go through several layers of inner pixel
detectors and to be reconstructed as (disappearing) tracks.  Such
disappearing tracks can be used not only for the reduction of standard
model backgrounds but also for the determination of the lifetime of
charged Wino.

Expecting that there exist several layers of pixel detectors, let
$\tilde{W}^\pm_i$ ($i=1$ $-$ $n_A$) be charged Winos which arrive
$A$-th layer of the pixel detector before decaying.  Here, $n_A$ is
the number of charged Wino samples available for the lifetime
measurement.  Then, the expectation value of the number of charged
Winos which arrive $B$-th layer ($B>A$) is
\begin{align}
  \langle n_B \rangle  (\tau) = \sum_{i=1}^{n_A} p_i (\tau),
\end{align}
with
\begin{align}
  p_i (\tau) \equiv
  e^{-(L_T^{(B)}-L_T^{(A)})/\tau \beta_i \gamma_i \sin\theta_i},
  \label{p_i}
\end{align}
where $L_T^{(A)}$ and $L_T^{(B)}$ are transverse distance from the
interaction point to the $A$- and $B$-th layers, respectively,
$\beta_i$ is the velocity of $i$-th Wino, $\gamma_i\equiv
1/\sqrt{1-\beta_i^2}$, and $\theta_i$ is the angle between the proton
beam and the direction of the momentum of $\tilde{W}^\pm_i$.  Notice
that $\langle n_B \rangle$ is an increasing function of $\tau$ and is
sensitive to the lifetime of charged Wino.  Thus, with the
measurements of the numbers of charged Winos arriving at $A$- and
$B$-th layers, we may acquire information about $\tau$, assuming that
the velocity and the propagation direction of charged Winos are
measurable.  In particular, once $n_B$ (i.e., the number of charged
Winos reaching to the $B$-th layer) is measured, the best-fit value of
the lifetime is given by solving
\begin{align}
  \langle n_B \rangle  (\tau^{\rm (best)}) = n_B.
\end{align}

The propagations of charged Winos from the $A$-th layer to the $B$-th
are multiples of Bernoulli processes with various probabilities.
Assuming a test lifetime $\tau^{\rm (test)}$, the probability to
realize a specific value of $n_B$ for a given data set is expressed as \cite{Chen:1997}
\begin{align}
  P (n_B;\tau^{\rm (test)}) &=
  \left[ \prod_{i=1}^{n_A} (1-p_i^{\rm (test)}) \right]
  \sum_{i_1<i_2<\cdots<i_{n_B}}
  \left[
    \frac{p_{i_1}^{\rm (test)}}{1-p_{i_1}^{\rm (test)}}
    \frac{p_{i_2}^{\rm (test)}}{1-p_{i_2}^{\rm (test)}} \cdots
    \frac{p_{i_{n_B}}^{\rm (test)}}{1-p_{i_{n_B}}^{\rm (test)}}
    \right], \notag \\
  &= \frac{1}{n_A+1}
  \sum_{l=0}^{n_A}
  e^{-2\pi i n_A l/(n_A+1)}
  \prod_{k=1}^{n_A}
  \left[ 1+(1-e^{2\pi i l/(n_A+1)} p_k^{\rm (test)}) \right].
  \label{probdist}
\end{align}
where $p_{i}^{\rm (test)}\equiv p_i(\tau^{\rm (test)})$ and the sum is
taken for all the possible sets of $\{i_1,i_2,\cdots,i_{n_B}\}$.
The second equality follows from \cite{Fernandez:2010}, which reduces the cost of numerical calculation.

Once charged Winos are observed at future collider experiments with
the measurements of their velocities and directions (as well as $n_A$
and $n_B$), we may constrain the lifetime of charged Wino.  In our
analysis, we define $\alpha\ \%$ ``confidence interval,'' which we
denote $\{n_B\}_\alpha$, as follows.  We define integers $I_1$,
$\cdots$, $I_{n_A}$ such that $P (I_1; \tau^{\rm
  (test)})\geq\cdots\geq P (I_{n_A}; \tau^{\rm (test)})$. Then, the
confidence interval, $\{n_B\}_\alpha\equiv\{I_1,\cdots,I_N\}$, is
defined as
\begin{align}
  \sum_{I\in\{n_B\}_\alpha} P (I; \tau^{\rm (test)})
  - P (I_{N}; \tau^{\rm (test)})
  <
  \alpha\ \%
  \leq
  \sum_{I\in\{n_B\}_\alpha} P (I; \tau^{\rm (test)}).
\end{align}
A test lifetime $\tau^{\rm (test)}$ is allowed (excluded) if observed
$n_B$ is inside (outside) of the confidence interval calculated with
$\tau^{\rm (test)}$.  In the next section, we discuss how well we can
determine the lifetime using MC analysis.

\section{Monte Carlo Analysis}
\label{sec:mc}
\setcounter{equation}{0}

In this section, we show that the determination of the lifetime of
charged Wino is really possible using MC analysis. For simplicity,
motivated by the minimal gravity mediation model based on anomaly
mediation, we concentrate on the model in which gauginos (i.e., Bino,
Wino, and gluino) are the only superparticles accessible with FCC-hh;
other superparticles are assumed to be too heavy to be produced.  In
addition, Winos are assumed to be lighter than Bino and gluino.  We
adopt three Sample Points which are based on the minimal gravity
mediation model \cite{Asai:2019wst}.  The Sample Points are shown in
Table \ref{table:samplept}; on the table, the masses of Bino, Wino,
and gluino (denoted as $m_{\tilde{B}}$, $m_{\tilde{W}}$, and
$m_{\tilde{g}}$, respectively) as well as the gluino pair production
cross section and the canonical luminosity in our analysis are given.
In our study, we take
$\mathrm{Br}(\tilde{g}\rightarrow\tilde{W}\bar{q}q)=\mathrm{Br}(\tilde{g}\rightarrow\tilde{B}\bar{q}q)=0.5$
(with $q$ being quarks), with the assumption of the flavor
universality of the final-state quarks.  For the cases of the gluino
mass of $6$ and $7$ TeV, we assume the integrated luminosity of ${\cal
  L}=10\ {\rm ab}^{-1}$, while ${\cal L}=30\ {\rm ab}^{-1}$ is used
for the case of $m_{\tilde{g}}=8\ {\rm TeV}$ to compensate the
smallness of the gluino production cross section.

\begin{table}[t]
  \begin{center}
    \begin{tabular}{c|ccc}
      \hline\hline
      & Point 1 & Point 2 & Point 3
      \\
      \hline
      $m_{\tilde{B}}$ [GeV] & $3660$ & $4060$ & $4470$
      \\
      $m_{\tilde{W}}$ [GeV] & $2900$ & $2900$ & $2900$
      \\
      $m_{\tilde{g}}$ [GeV] & $6000$ & $7000$ & $8000$
      \\
      $\sigma (pp\rightarrow \tilde{g}\tilde{g})$ [fb]
      & $7.9$ & $2.7$ & $1.0$
      \\
      ${\cal L}$ [1/ab]
      & $10$ & $10$ & $30$
      \\
      \hline\hline
    \end{tabular}
    \caption{Gaugino masses and the gluino pair production cross
      section (for the center-of-mass of $100\ {\rm TeV}$) for the
      Sample Points 1, 2, and 3.  We also show the canonical
      luminosities used for the analysis for these Sample Points.}
    \label{table:samplept}
  \end{center}
\end{table}

Our method of event generation is mostly the same as that adopted in
\cite{Asai:2019wst}.  We use {\tt MadGraph5\_aMC@NLO}
\cite{Alwall:2011uj, Alwall:2014hca} for the generation of
$pp\rightarrow\tilde{g}\tilde{g}$ and
$pp\rightarrow\tilde{W}^{+}\tilde{W}^{-}+\mathrm{jets}$ events.  The
results are passed to {\tt PYTHIA8} \cite{Sjostrand:2014zea} for the
decay and hadronization processes.  Then, {\tt Delphes} ({\tt v3.4.1})
\cite{deFavereau:2013fsa} is used for a fast detector simulation; we
use the card {\tt FCChh.tcl} included in the package.  The velocities
of charged Winos are smeared by our original code.  We expect that, at
FCC-hh, the charged Wino track can be reconstructed and that
information about the time of flight is available if the charged Wino
hits several layers of pixel detector.  We assume $6\ \%$ error in the
velocity measurement \cite{Asai:2019wst}; for reconstructed charged
Wino tracks, the observed values of the Wino velocity are determined
as follows:
\begin{align}
  \beta = (1 + 6\ \% \times Z) \beta^{\rm (true)},
  \label{betaerror}
\end{align}
where $\beta$ and $\beta^{\rm (true)}$ are observed and true values of
the velocity and $Z$ is the $(0,1)$ Gaussian random variable.  We
neglect the error in the measurement of the directions of Wino tracks.

\begin{figure}[t]
  \begin{center}
    \includegraphics[width=0.65\textwidth]{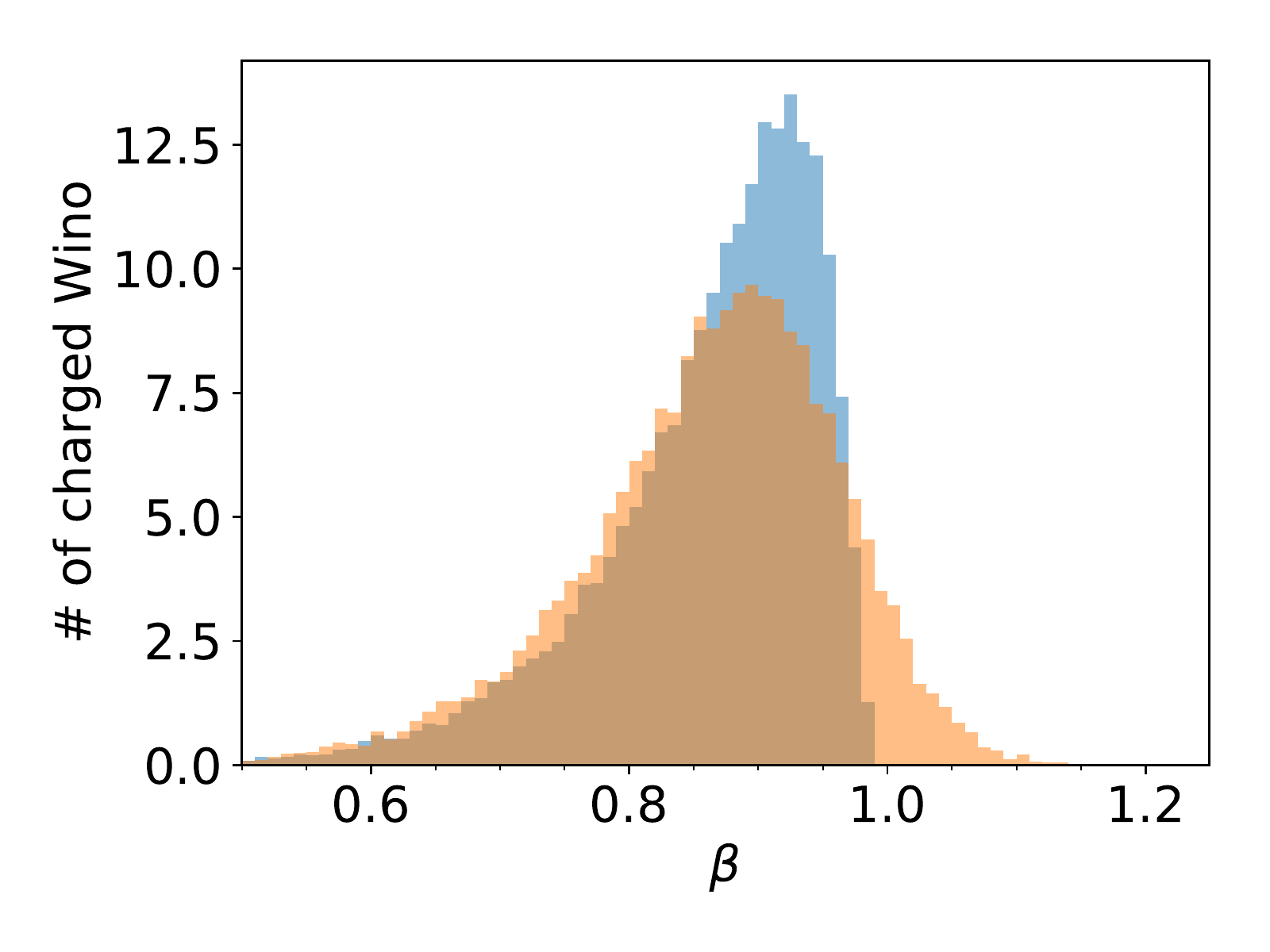}
    \caption{\small Distributions of the true (blue) and observed
      (orange) velocities of charged Winos for the Sample Point 1.
      The height of the histogram is the number of charged Winos in
      the bin for events satisfying the Requirements 1 and 2.  The bin
      width is $\Delta\beta=0.01$. }
    \label{fig:beta}
  \end{center}
\end{figure}

In the following analysis, we assume that the charged Winos are
required to hit at least inner four layers of the pixel detector for
the track reconstruction (and also for background reduction).  In
order to eliminate standard model backgrounds, we use only the events
satisfying the following requirements:
\begin{itemize}
\item[1.] There exist two ``long enough'' Wino-like tracks.  The
  transverse length of the tracks should be longer than the transverse
  distance to the 4th pixel detector $L_{T}^{(4)}$.  In addition, the
  pseudo-rapidities ($\eta$) of the tracks should satisfy
  $|\eta|<1.5$.
\item[2.] The missing transverse energy (MET) should be larger than
  $1\ {\rm TeV}$.
\end{itemize}
With these requirements, we expect that the standard model backgrounds
can become negligible, as discussed in \cite{Asai:2019wst}.

In Fig.\ \ref{fig:beta}, we show the distribution of $\beta$ and
$\beta^{\rm (true)}$ for the Sample Point 1, using the events
satisfying Requirements 1 and 2.  (Here, 100 sets of the event samples
are used for the figure.)  We can see that the peak of the
distribution is shifted to the small value of the velocity after the
smearing.  We will discuss its effects of the lifetime determination
later.

For the determination of $\tau$, we use charged Winos in the events
satisfying the requirements given above, and hence we take
$(A,B)=(4,5)$.  In addition, for the lifetime determination, velocity
information about charged Winos is necessary.  In order for a good
velocity measurement, we use only charged Winos whose (observed)
velocity is smaller than $0.85$ for the lifetime determination.

\begin{figure}[t]
  \begin{center}
    \includegraphics[width=0.65\textwidth]{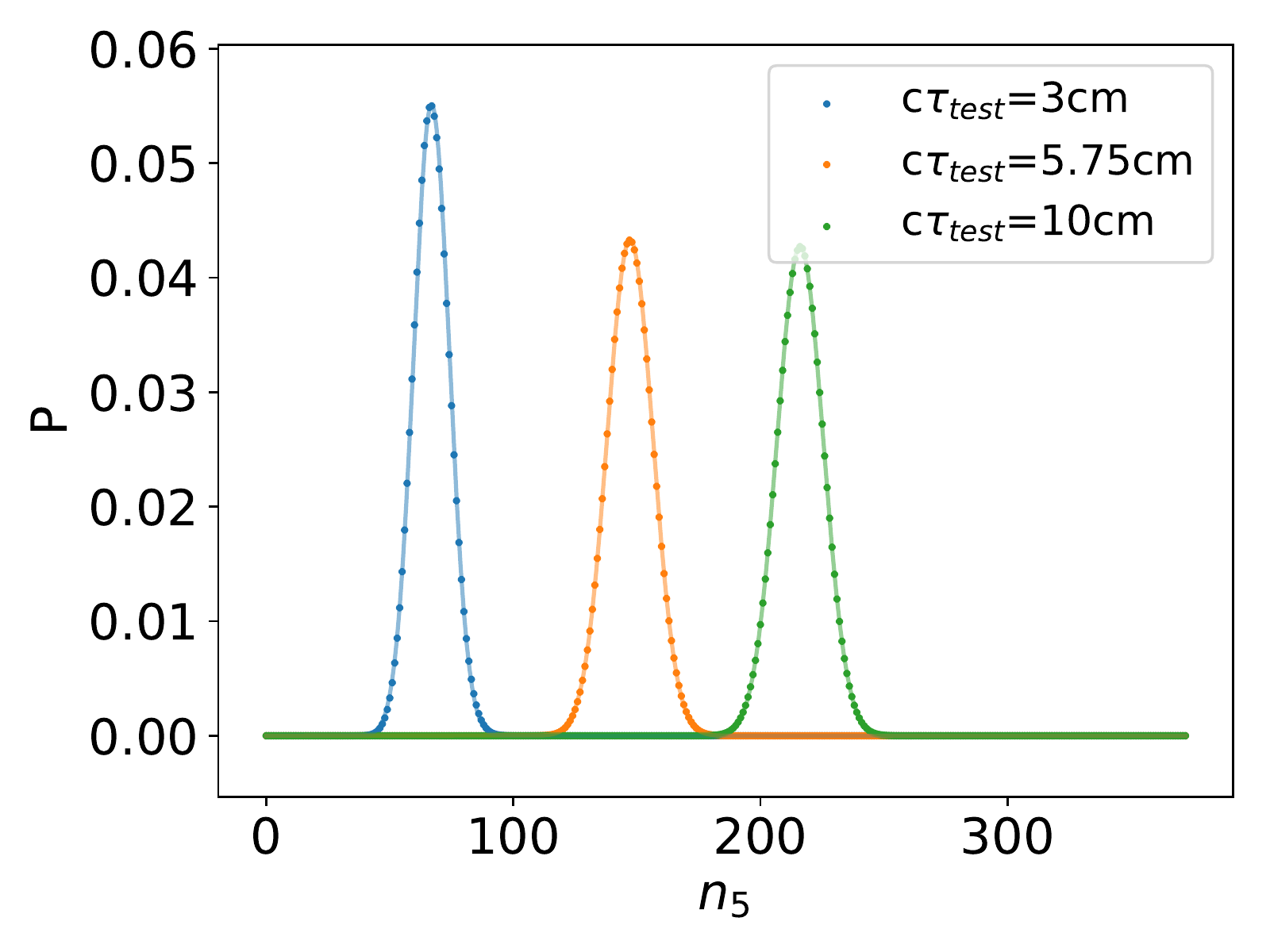}
    \caption{\small The dots show $P (n_5; \tau^{\rm (test)})$ for
      $L_{T}^{(4)}=10\ {\rm cm}$ and $L_{T}^{(5)}=15\ {\rm cm}$, using
      the event samples generated from the Sample Point 1 with
      $c\tau=5.75\ {\rm cm}$.  $c\tau^{\rm (test)}$ is taken to be
      $3\ {\rm cm}$ (blue), $5.75\ {\rm cm}$ (red), and
      $10\ {\rm cm}$ (green) from left to right.
      The solid lines show the Gaussian distribution
      $N(\hat{n}_5(\tau^{\rm (test)}),\hat{n}_5(\tau^{\rm (test)}))$,
      where $\hat{n}_5(\tau^{\rm (test)})$ is the value of $n_5$ which
      maximizes $P (n_5; \tau^{\rm (test)})$.}
    \label{fig:probdist}
  \end{center}
\end{figure}

First, we consider the case where the gluino mass is light enough so
that the gluino pair production dominates the SUSY events at FCC-hh
(i.e., the case of the Sample Points 1 $-$ 3).  In
Fig.\ \ref{fig:probdist}, taking $L_{T}^{(4)}=10\ {\rm cm}$ and
$L_{T}^{(5)}=15\ {\rm cm}$, we plot $P (n_5; \tau^{\rm (test)})$
taking $c\tau^{\rm (test)}=3\ {\rm cm}$ (blue), $c\tau^{\rm
  (test)}=5.75\ {\rm cm}$ (red), and $c\tau^{\rm (test)}=10\ {\rm cm}$
(green), using the event sample generated from the Sample Point 1 with
$c\tau=5.75\ {\rm cm}$.  We can see that the behavior of $P (n_5;
\tau^{\rm (test)})$ is strongly dependent on $\tau^{\rm (test)}$.
Thus, with the measurement of the number of charged Winos reaching to
the 5th layer, we can obtain information about the lifetime.  We also
plot the Gaussian distribution $N(\hat{n}_5(\tau^{\rm
  (test)}),\hat{n}_5(\tau^{\rm (test)}))$, where $\hat{n}_5(\tau^{\rm
  (test)})$ is the value of $n_5$ which maximizes $P (n_5; \tau^{\rm
  (test)})$.  We can see that the probability distribution is well
approximated by the Gaussian distribution when the number of charged
Winos reaching to the 5th layer is large enough.

\begin{figure}
  \begin{center}
    \includegraphics[width=0.65\textwidth]{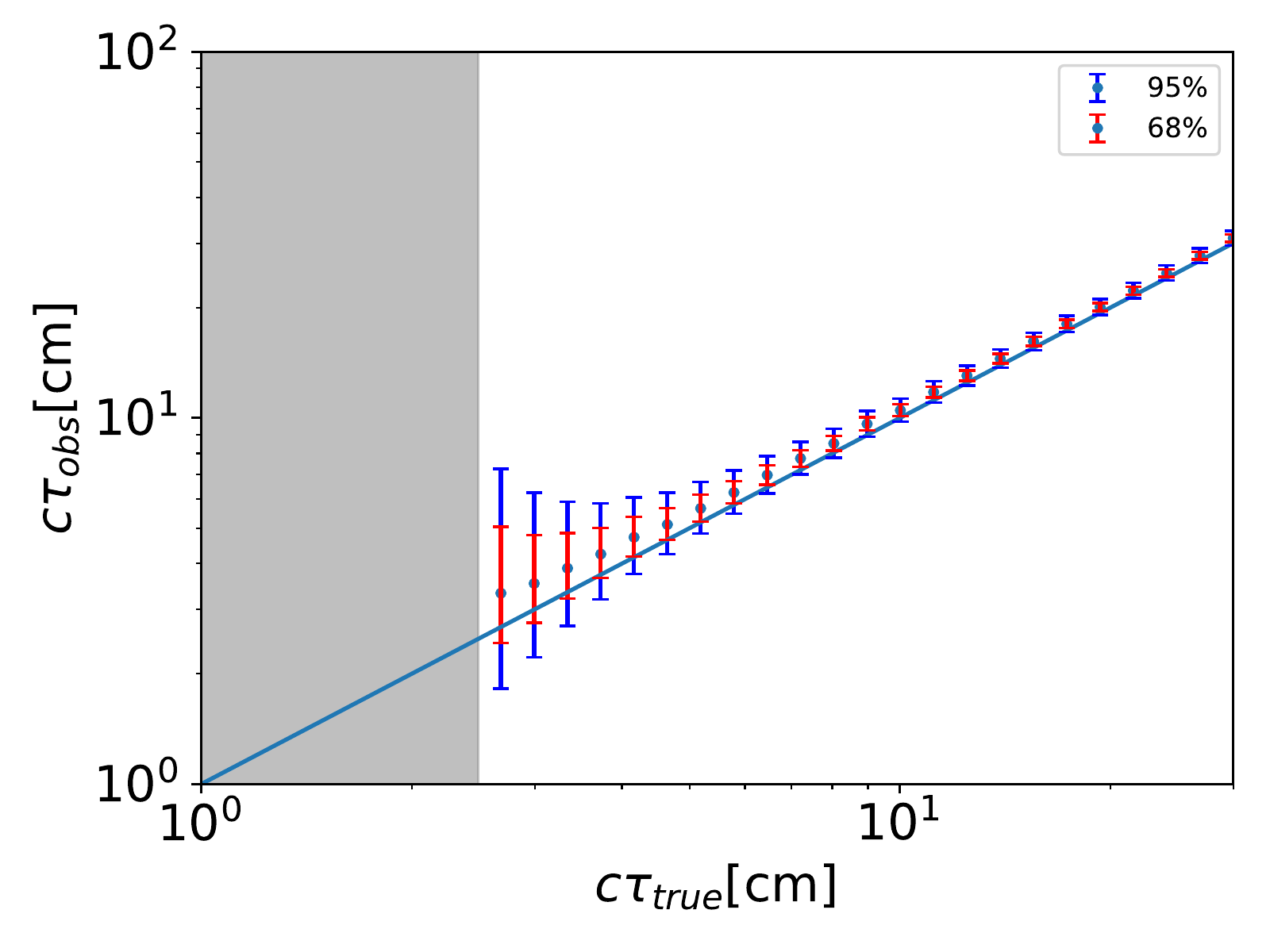}
    \caption{\small Expected $68\, \%$ and $95\, \%$ confidence level
      lower and upper bounds from the lifetime measurements, as well
      as the best-fit values, for the Sample Point 1.}
    \label{fig:pp2gogo6TeV}
  \end{center}
  \begin{center}
    \includegraphics[width=0.65\textwidth]{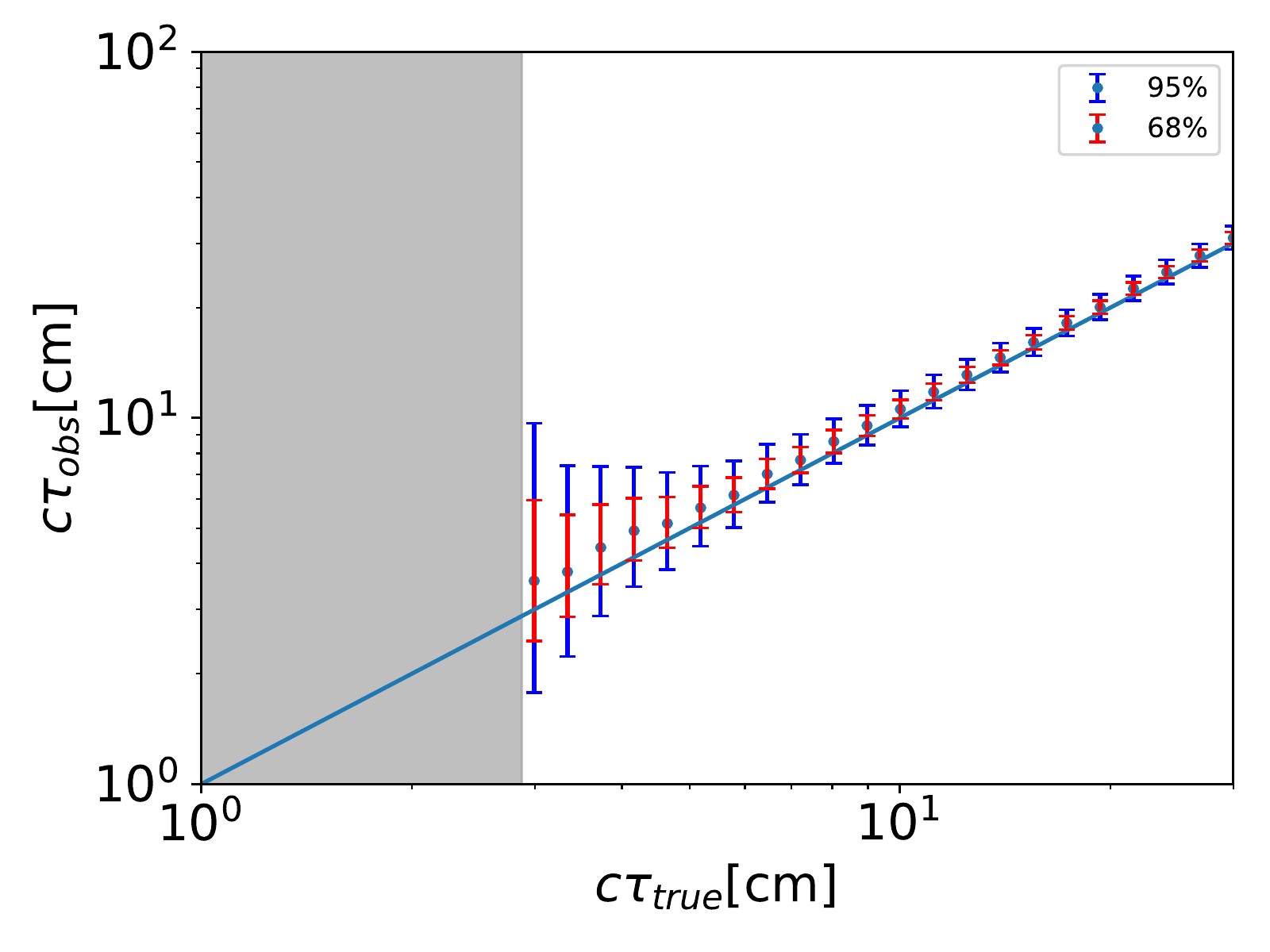}
    \caption{\small Same as Fig.\ \ref{fig:pp2gogo6TeV}, except for
      the Sample Point 2.}
    \label{fig:pp2gogo7TeV}
  \end{center}
\end{figure}

\begin{figure}[t]
  \begin{center}
    \includegraphics[width=0.65\textwidth]{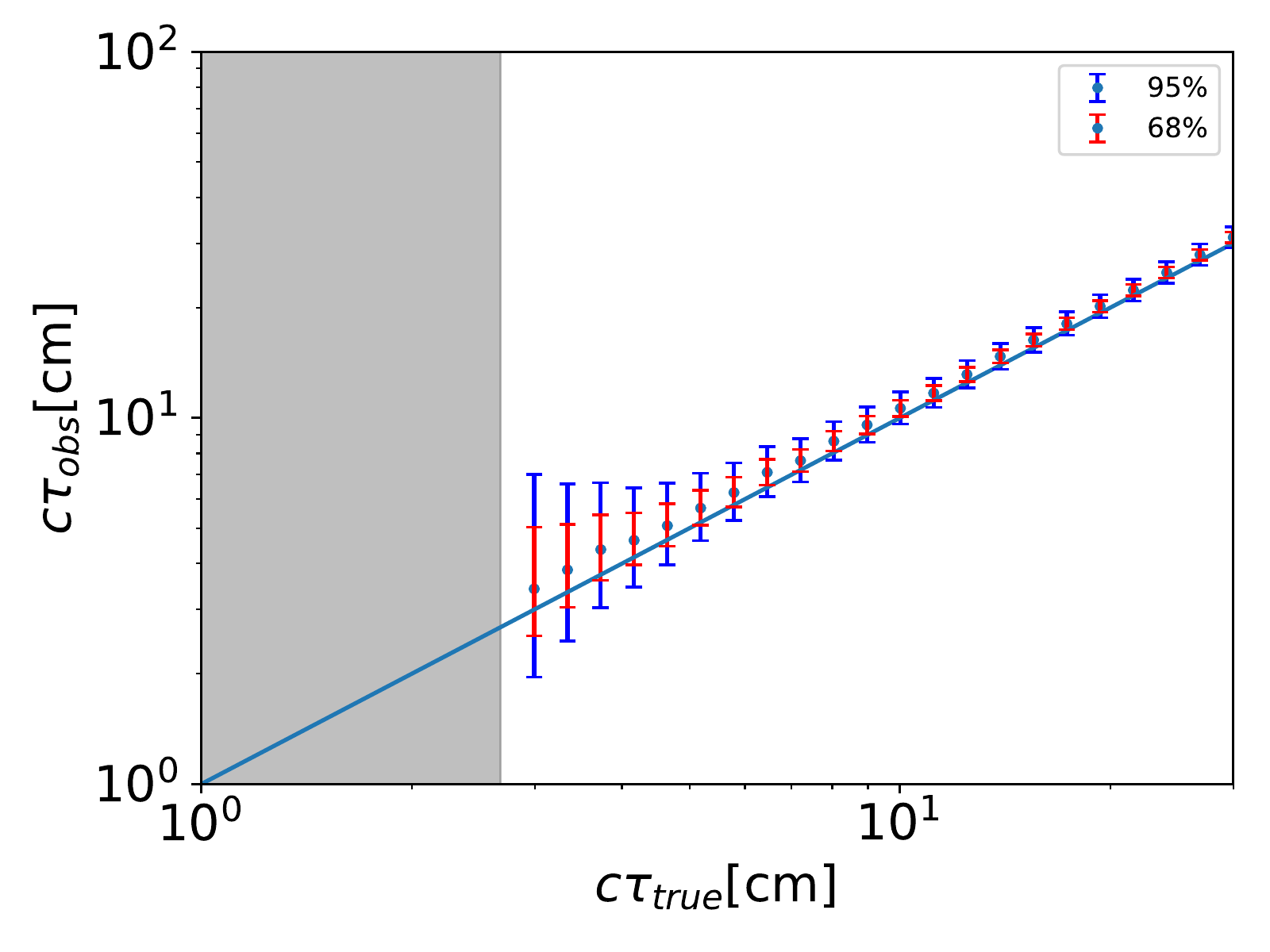}
    \caption{\small Same as Fig.\ \ref{fig:pp2gogo6TeV}, except for
      the Sample Point 3.}
    \label{fig:pp2gogo8TeV}
  \end{center}
\end{figure}

In Figs.\ \ref{fig:pp2gogo6TeV} $-$ \ref{fig:pp2gogo8TeV}, we plot the
best-fit value of the lifetime as well as the expected lower and upper
bounds for various choices of $\tau$ for the Sample Points 1 $-$ 3.
Here, we use 100 independent data sets for each Sample Point, and
determine the lower and upper bounds on the lifetime for each data set
using the probability distribution defined in Eq.\
\eqref{probdist}. The expected lower and upper bounds shown in the
figures are obtained by taking the median of those from 100 independent data
sets.  The regions with $\tau$ giving rise to $\langle n_4\rangle<10$,
for which our method of the lifetime measurement becomes difficult,
are shaded.

We can see that the best-fit values of the lifetime are systematically
overestimated in the present analysis.  This is mainly due to the
error in the velocity measurement.  Here, the Wino velocity is assumed
to be measured with the $6\ \%$ accuracy (see Eq.\ \eqref{betaerror}).
As one can see in Fig.\ \ref{fig:beta}, the observed Wino velocity is
likely to be smaller than the true value, resulting in the
overestimation of the lifetime.  We checked that the best-fit values
become consistent with the input values if the true value of the Wino
velocity is used in the analysis. We assume that such a systematic
effect originating from the velocity measurement can be well understood
in the actual experiment.  Thus, we will not include the shift of
the best-fit value in estimating the uncertainty of the lifetime
determination.

\begin{figure}[t]
  \begin{center}
    \includegraphics[width=0.65\textwidth]{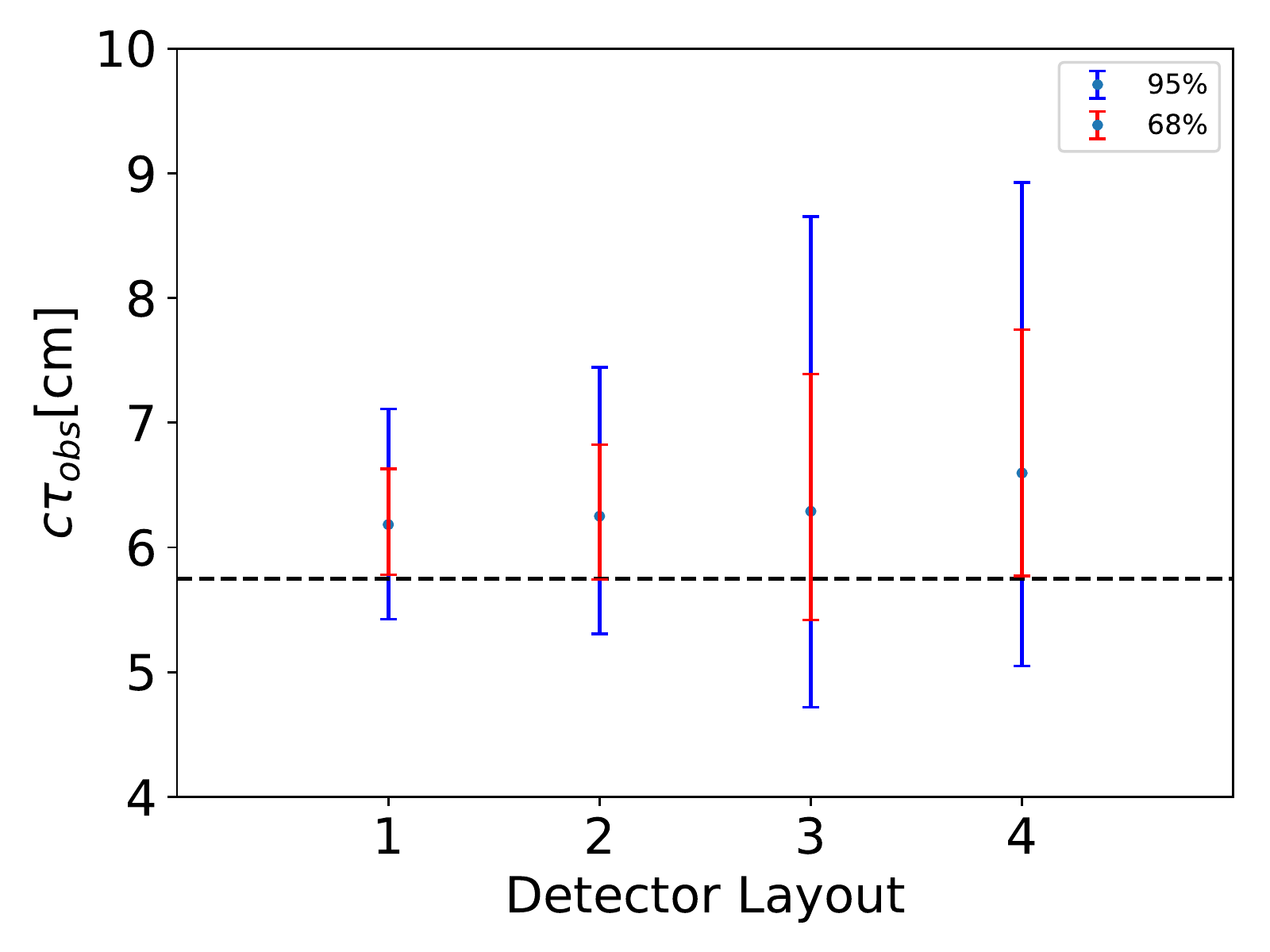}
    \caption{\small Accuracies of the lifetime determination for
      several choices of the detector layout for the Sample Point 1
      with $c\tau=5.75\ {\rm cm}$.  Here, we take $(L_{T}^{(4)},
      L_{T}^{(5)})=(10\ {\rm cm}, 15\ {\rm cm})$, $(11\ {\rm cm},
      15\ {\rm cm})$, $(15\ {\rm cm}, 20\ {\rm cm})$, and $(15\ {\rm cm},
       27\ {\rm cm})$ from left to right.}
    \label{fig:pp2gogo6TeVLayout}
  \end{center}
\end{figure}

Next, we show how the accuracy of the lifetime determination depends
on the detector layouts.  The accuracy depends on the distances to the
pixel layers.  In Fig.\ \ref{fig:pp2gogo6TeVLayout}, for the Sample
Point 1 with taking $c\tau=5.75\ {\rm cm}$, we show the expected lower
and upper bounds for several choices of the distances to the layers.
As expected, the accuracy becomes worse as the distances to the 4th
layer becomes longer; with larger $L_{T}^{(4)}$, the number of charged
Winos available for the analysis becomes smaller.  

We define the uncertainties in the lifetime determination as
\begin{align}
  \delta \tau^{(\pm)} \equiv | \tau^{(\pm)} - \tau^{\rm (best)} |,
\end{align}
where $\tau^{(-)}$ and $\tau^{(+)}$ are lower and upper bounds on the
lifetime for a given confidence interval, respectively.  For the
Sample Points $1$ $-$ $3$ with $c\tau=5.75\ {\rm cm}$, the values of
$\tau^{(\pm)}$ for several choices of detector layouts are summarized
in Table \ref{table:error}.  In the same Table, we show the median
values of $n_4$ and $n_5$ for each detector layout.  For some sample
points, $n_4$ for $(L_{T}^{(4)}, L_{T}^{(5)})=(15\ {\rm cm}, 20\ {\rm
  cm})$ and that for $(15\ {\rm cm}, 27\ {\rm cm})$ slightly differ;
it is due to  statistical fluctuations.

\begin{table}
  \begin{center}
    \begin{tabular}{c|cccccc}
      \multicolumn{5}{l}
      {Sample Point 1}
      \\
      \hline\hline
      $(L_{T}^{(4)},L_{T}^{(5)})$
      & $c\delta\tau^{(-)}$ ($68\, \%$) & $c\delta\tau^{(+)}$ ($68\, \%$)
      & $c\delta\tau^{(-)}$ ($95\, \%$) & $c\delta\tau^{(+)}$ ($95\, \%$)
      & $n_4$ & $n_5$
      \\
      \hline
      $(10\ {\rm cm}, 15\ {\rm cm})$
      & 0.40 & 0.45
      & 0.76 & 0.93
      & 400 & 180
      \\
      $(11\ {\rm cm}, 15\ {\rm cm})$
      & 0.51 & 0.57
      & 0.94 & 1.2
      & 290 & 150
      \\
      $(15\ {\rm cm}, 20\ {\rm cm})$
      & 0.87 & 1.1
      & 1.6 & 2.4
      & 85 & 40
      \\
      $(15\ {\rm cm}, 27\ {\rm cm})$
      & 0.83 & 1.2
      & 1.5 & 2.3
      & 85 & 16
      \\
      \hline\hline
    \end{tabular}

    \begin{tabular}{c|cccccc}
      \multicolumn{5}{l} {}
      \\
      \multicolumn{5}{l}
      {Sample Point 2}
      \\
      \hline\hline
      $(L_{T}^{(4)},L_{T}^{(5)})$
      & $c\delta\tau^{(-)}$ ($68\, \%$) & $c\delta\tau^{(+)}$ ($68\, \%$)
      & $c\delta\tau^{(-)}$ ($95\, \%$) & $c\delta\tau^{(+)}$ ($95\, \%$)
      & $n_4$ & $n_5$
      \\
      \hline
      $(10\ {\rm cm}, 15\ {\rm cm})$
      & 0.65 & 0.74
      & 1.2 & 1.5
      & 170 & 78
      \\
      $(11\ {\rm cm}, 15\ {\rm cm})$
      & 0.80 & 0.90
      & 1.4 & 2.0
      & 120 & 66
      \\
      $(15\ {\rm cm}, 20\ {\rm cm})$
      & 1.4 & 1.9
      & 2.3 & 4.2
      & 38 & 17
      \\
      $(15\ {\rm cm}, 27\ {\rm cm})$
      & 1.2 & 1.7
      & 2.1 & 3.5
      & 40 & 7
      \\
      \hline\hline
    \end{tabular}

    \begin{tabular}{c|cccccc}
      \multicolumn{5}{l} {}
      \\
      \multicolumn{5}{l}
      {Sample Point 3}
      \\
      \hline\hline
      $(L_{T}^{(4)},L_{T}^{(5)})$
      & $c\delta\tau^{(-)}$ ($68\, \%$) & $c\delta\tau^{(+)}$ ($68\, \%$)
      & $c\delta\tau^{(-)}$ ($95\, \%$) & $c\delta\tau^{(+)}$ ($95\, \%$)
      & $n_4$ & $n_5$
      \\
      \hline
      $(10\ {\rm cm}, 15\ {\rm cm})$
      & 0.56 & 0.64
      & 1.0 & 1.3
      & 230 & 100
      \\
      $(11\ {\rm cm}, 15\ {\rm cm})$
      & 0.71 & 0.79
      & 1.2 & 1.6
      & 170 & 90
      \\
      $(15\ {\rm cm}, 20\ {\rm cm})$
      & 1.1 & 1.4
      & 2.0 & 3.2
      & 53 & 25
      \\
      $(15\ {\rm cm}, 27\ {\rm cm})$
      & 1.0 & 1.5
      & 1.9 & 3.3
      & 53 & 10
      \\
      \hline\hline
    \end{tabular}

    \begin{tabular}{c|cccccc}
      \multicolumn{5}{l} {}
      \\
      \multicolumn{5}{l}
      {$pp\rightarrow\tilde{W}^{+}\tilde{W}^{-}+\mathrm{jets}$}
      \\
      \hline\hline
      $(L_{T}^{(4)},L_{T}^{(5)})$
      & $c\delta\tau^{(-)}$ ($68\, \%$) & $c\delta\tau^{(+)}$ ($68\, \%$)
      & $c\delta\tau^{(-)}$ ($95\, \%$) & $c\delta\tau^{(+)}$ ($95\, \%$)
      & $n_4$ & $n_5$
      \\
      \hline
      $(10\ {\rm cm}, 15\ {\rm cm})$
      & 1.1 & 1.5
      & 1.9 & 3.2
      & 60 & 28
      \\
      $(11\ {\rm cm}, 15\ {\rm cm})$
      & 1.3 & 1.8
      & 2.2 & 4.0
      & 45 & 24
      \\
      $(15\ {\rm cm}, 20\ {\rm cm})$
      & 2.2 & 3.6
      & 3.4 & 9.2
      & 15 & 7
      \\
      $(15\ {\rm cm}, 27\ {\rm cm})$
      & 1.8 & 3.8
      & 3.2 & 8.1
      & 15 & 3
      \\
      \hline\hline
    \end{tabular}

    \caption{The expected uncertainties in the lifetime determination
      in units of cm, adopting several choices of detector layouts.
      The input value of the lifetime is taken to be $c\tau=5.75\ {\rm
        cm}$.  The median values of $n_4$ and $n_5$ are also shown.}
    \label{table:error}
  \end{center}
\end{table}

\begin{figure}[t]
  \begin{center}
    \includegraphics[width=0.65\textwidth]{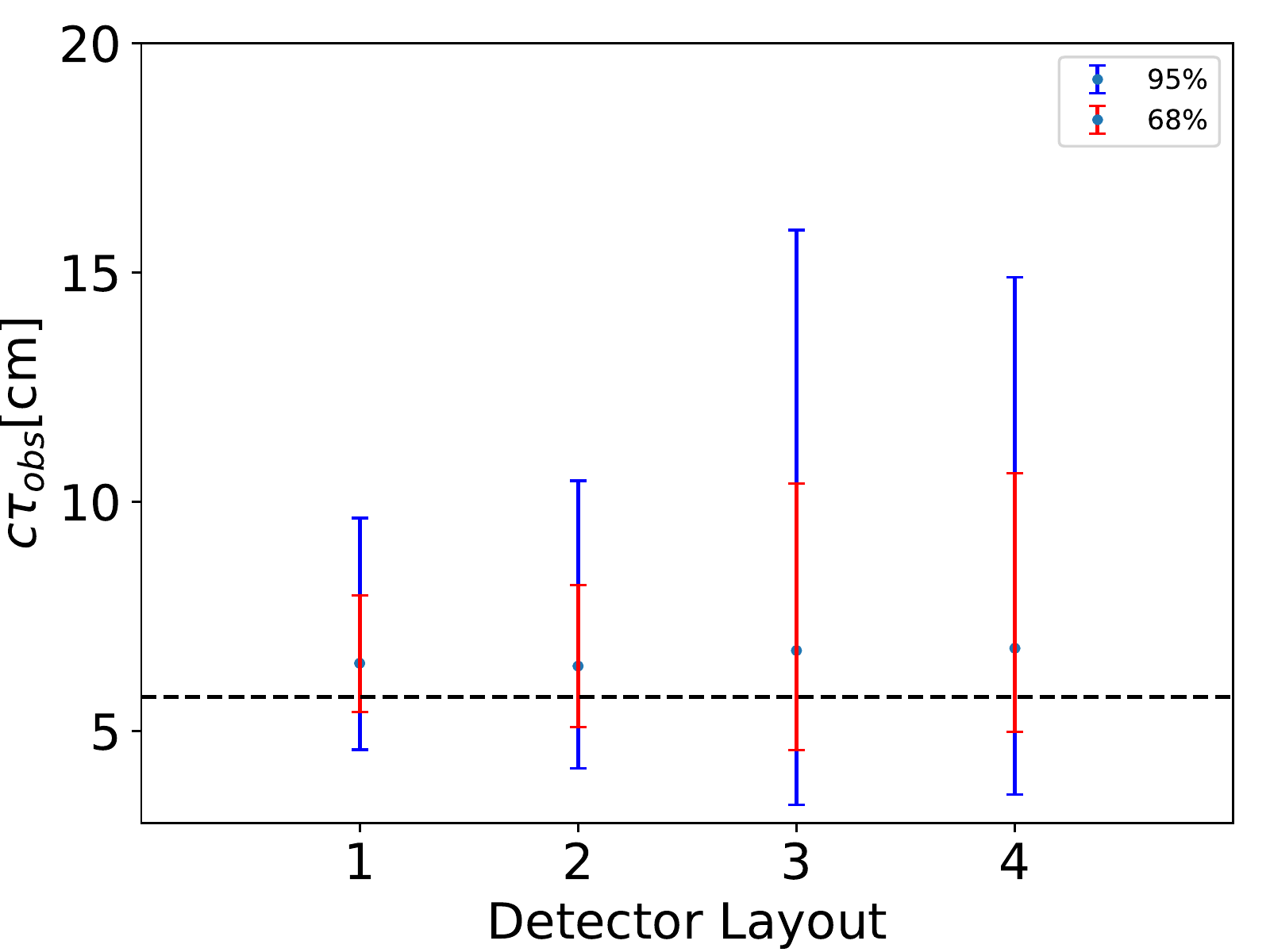}
    \caption{\small Accuracies of the lifetime determination for
      several choices of the detector layout for
      $m_{\tilde{W}}=2.9\ {\rm TeV}$ and $c\tau=5.75\ {\rm cm}$, using
      the process
      $pp\rightarrow\tilde{W}^{+}\tilde{W}^{-}+\mathrm{jets}$.  Here,
      we take $(L_{T}^{(4)}, L_{T}^{(5)})=(10\ {\rm cm}, 15\ {\rm
        cm})$, $(11\ {\rm cm}, 15\ {\rm cm})$, $(15\ {\rm cm},
      20\ {\rm cm})$, and $(15\ {\rm cm}, 27\ {\rm cm})$ from left to
      right.}
    \label{fig:DYjLayout}
  \end{center}
\end{figure}

When gluino is too heavy to be produced, the gluino pair production
process cannot be used for our analysis.  Even in such a case, we may
use the direct production of Winos for the lifetime determination.  In
particular, if the Wino mass is $\sim 2.9\ {\rm TeV}$, which is the
value of the Wino mass relevant for the thermal Wino dark matter
scenario, charged Wino can be within the discovery reach of the
disappearing track search at the FCC-hh \cite{Saito:2019rtg}.  This
fact indicates that the lifetime determination is also possible.  In
order to see how well we can determine the lifetime, we consider the
process $pp\rightarrow\tilde{W}^{+}\tilde{W}^{-}+\mathrm{jets}$.
Here, the extra jets are required for the trigger selection (as well
as for the kinematical cut of our choice).  For the events, we impose
the Requirements 1 and 2, which we mentioned before.  Then, using the
events satisfying the Requirements, we determine the best-fit value of
the lifetime as well as the confidence interval.  In Fig.\
\ref{fig:DYjLayout}, with adopting several detector layouts, we show
the expected accuracy of the determination of the Wino lifetime,
taking $m_{\tilde{W}}=2.9\ {\rm TeV}$, $c\tau=5.75\ {\rm cm}$, and
${\cal L}=30\ {\rm ab}^{-1}$. Here, we use independent 500 data sets
to calculate the median values of best-fit lifetime as well as lower
and upper bounds. However $n_5 = 0$ in 5 data sets of them, so we
exclude these sets before taking the median.  The uncertainties for
our choices of detector layouts are also summarized in Table
\ref{table:error}, taking $c\tau=5.75\ {\rm cm}$.  One can see that
the uncertainties are larger than the cases of the gluino pair
production events.  This is mainly due to the smallness of the cross
section for the
$pp\rightarrow\tilde{W}^{+}\tilde{W}^{-}+\mathrm{jets}$ process.  Even
so, the Wino lifetime can be determined with a relatively good
accuracy, i.e., $\delta\tau^{(\pm)}/\tau\sim O(10)\ \%$, in particular
if a compact pixel detector with $L_T^{(5)}\sim 15\ {\rm cm}$ is
available.  We also comment here that, for $L_{T}^{(4)}=15\ {\rm cm}$,
the layout with $L_{T}^{(5)}=27\ {\rm cm}$ gives better accuracy than
$L_{T}^{(5)}=20\ {\rm cm}$.  This is due to the fact that the accuracy
becomes worse when $L_{T}^{(A)}$ and $L_{T}^{(B)}$ take too close
values.

\section{Conclusions and Discussion}
\label{sec:conclusion}
\setcounter{equation}{0}

In this letter, we have discussed the possibility to determine the
lifetime of charged Wino in supersymmetric model, assuming that the
neutral Wino is the LSP.  In such a case, the lifetime of the charged
Wino is given by $c\tau\simeq 5.75\ {\rm cm}$, for which we have seen
that a significant number of charged Winos may hit several layers of
inner pixel detector and may be used for the lifetime determination.
Concentrating on the case with the Wino mass of $2.9\ {\rm TeV}$, which
is the relevant value to make thermal relic Wino as dark matter, we have
studied the prospect of the Wino lifetime determination at FCC-hh.

If gluino is within the kinematical reach of FCC-hh, we may use the
charged Winos produced by the decay of gluino.  In such a case, the
Wino lifetime may be determined with the accuracy of $14\ \%$
($30\ \%$) for the $68\ \%$ ($95\ \%$) confidence interval.  Even if
gluino is out of the kinematical reach, we may use the charged Winos
produced by the process
$pp\rightarrow\tilde{W}^{+}\tilde{W}^{-}+\mathrm{jets}$.  In such a
case, the accuracy of the lifetime determination becomes worse, but
still it can be $43\ \%$ ($92\ \%$) for the $68\ \%$ ($95\ \%$)
confidence interval.  These measurements of the Wino lifetime provides
an important confirmation that the observed charged particle is really
$\tilde{W}^\pm$.

Finally, we comment on the effects of the accidental alignments of the
hits on the pixel detector, which has been neglected in our analysis.
Potentially, the most serious effect may come from the hits on the 5th
layer near the trajectories of the true charged Wino tracks.  If there
exist such hits for charged Winos which decay before reaching 5th
layer, they affect the measurement of $n_5$.  According to the study
of the fake tracks given in \cite{Saito:2019rtg}, however, the
probability to have fake charged Wino tracks decreases by a factor of
$O(100)$ with requiring a hit on an extra layer.  Thus, we estimate
that the error of $n_5$ due to the accidental alignment is less than
$O(10^{-2})\times n_4$, which is negligible in the situation of our
study.

\section*{Acknowledgments}
This work was supported by MEXT KAKENHI Grant number JP16K21730 and
JSPS KAKENHI Grant (Nos.\ 17J00813 [SC], 16H06490 [TM], 18K03608 [TM],
and 18J11405 [MS]).

\end{document}